\documentclass[5p]{elsarticle}

\usepackage{graphicx}
\usepackage{bm}
\usepackage{lineno}
\usepackage{lineno, blindtext}

\begin{document} 
\biboptions{sort&compress}


\title{Effect of valence fluctuations on the  ground state properties of SmB$_6$}

\author{Marianna Batkova\corref{Corresponding author}}
\ead{batkova@saske.sk}
\author{Ivan Batko}
\address{Institute of Experimental Physics,
 Slovak   Academy  of Sciences, Watsonova 47,
 040~01~Ko\v {s}ice, Slovakia}

\date{\today}

\begin{abstract}     

We argue that because of valence-fluctuation caused dynamical changes (fluctuations) of impurity energies in the impurity band of SmB$_6$, 
energies of electrons occupying impurity sites can be due to the uncertainty principle only estimated with corresponding uncertainty,   
so they can be represented by energy intervals of non-zero width.
As a consequence, both occupied as well as unoccupied states can be found in the impurity band 
above as well as below Fermi level even in the ground state. 
Therefore the subsystem of localized charge carriers in the ground state of SmB$_6$ cannot be described 
by an energy distribution function expected for $T=0$~K. This fundamental conclusion adds another reason 
for absence of resistivity divergence in SmB$_6$ at lowest temperatures, 
and sheds new light on interpretation of experimental data obtained for SmB$_6$ and similar systems 
at lowest temperatures.
\end{abstract}%

\maketitle


SmB$_6$  is a prototypical mixed valence material revealing properties of a narrow-gap semiconductor 
down to temperatures of a few Kelvin  \cite{Allen79,Kasuya79, Cooley95,Batkova06}.
Paradoxically, at lowest temperatures it   
 exhibits metallic-like conductivity, which is moreover smaller 
 than the Mott's minimum metallic conductivity \cite{Allen79,Kasuya79, Cooley95}.
Such temperature non-activated electrical transport of SmB$_6$ at lowest temperatures
cannot be attributed to any scattering scenario known for metals, 
%
because extremely high value of the residual resistivity  
would unconditionally require superunitary scattering 
with unphysical concentration of scattering centers \cite{Allen79,Kasuya79, Cooley95}, at least 80 per unit cell \cite{Cooley95}. 
This fundamental discrepancy between theory and experiment 
is a subject of long-standing controversy precluding to conclude whether 
ground state of SmB$_6$ is metallic, or insulating.
In fact, 
 according to Mott-Ioffe-Regel viewpoint conventional Boltzmann transport theory 
becomes meaningless if the characteristic mean free path 
of the itinerant conduction electrons becomes comparable to, 
or less than the interatomic spacing \cite{Ioffe60,Mott71,Edwards98,Shklovskij84}.  
Therefore the requirement for superunitary scattering implicates that  		
		either (i) electrical conductivity is not homogeneous in the volume, i.e. material contains metallic regions forming a conductive path along the sample
		that is responsible for electrical conductivity at lowest temperatures  \cite{Dzero10,Dzero12,Lu13,Alexandrov13,Zhu13} 
		or (ii) electrical transport at lowest temperatures is realized via a hopping process,
		which however, has to be temperature non-activated  \cite{BaBa14}. 
The first mentioned approach is represented by scenarios supposing  
	 metallic surface in SmB$_6$ 
 that could be either of topological nature \cite{Dzero10,Dzero12,Lu13,Alexandrov13}
or due to "trivial" polarity-driven surface states \cite{Zhu13}. 
Several experimental observations indicate metallic surface transport in SmB$_6$ 
\cite{Wolgast13,Kim13,Syers15} 
while there exists huge research effort to prove the existence  
of topologically protected surface states \cite{Jiang13,Neupane13,Kim14,Li14,Ruan14,Phelan14,Nakajima15,Chen15}.  
On the other hand, latest developments provide also evidence of trivial surface states in SmB$_6$ \cite{Hlawenka15}.
It has been also suggested that only the most stoichiometric SmB$_6$ samples possess
a bulk gap necessary for the topological Kondo insulator state \cite{Valentine16}. 
However, up to now, convincing conclusion about the nature of metallic surface states in SmB$_6$ is still missing. 
The second mentioned approach is represented by a recently proposed model 
of valence-fluctuation induced hopping transport \cite{BaBa14}. 
The model is based on the fact that valence fluctuations (VFs) of Sm ions are {\em intrinsically}
accompanied by fluctuations of charge, ionic radii and magnetic moments of Sm ions,
what unconditionally causes fluctuations of the energies of impurities in the impurity band of SmB$_6$.
As a consequence there are created favorable conditions for
temperature non-activated hops (as in more details summarized below), 
resulting in temperature non-activated hopping transport \cite{BaBa14}.  
The model intrinsically infers an \emph{enhanced} conductivity of the surface layer 
because of higher concentration of lattice imperfections \cite{BaBa14},
thus resembling a characteristic property of topological insulators \cite{BaBa14}. 
In association with this it should be emphasized here that although the scenario 
as presented in our previous work \cite{BaBa14}  has a capability to explain properties of SmB$_6$ 
without necessity to consider Kondo topological insulator (KTI) scenario, it does not exclude a possibility of existence of topologically protected surface states in SmB$_6$. 
In fact, both of them can coexist, thus their features can be "mixed" in SmB$_6$.   

The purpose of this work is to point out that besides valence-fluctuation induced hopping transport as proposed before \cite{BaBa14}, 
another consequence of dynamical changes of the energies 
		of impurities in SmB$_6$ due to VFs is that the energy distribution function of the charge carrier subsystem
		in the ground state corresponds to presence of occupied states above Fermi level and empty ones below Fermi level, 
		thus	resembling effect of thermal broadening.
This implicates that temperature of charge carrier subsystem in the ground state of SmB${_6}$ 
cannot be $ T = 0$~K.

For convenience of further discussion
let us first recapitulate essential ideas of the original model of valence-fluctuation induced hopping transport \cite{BaBa14}. 
In a semiconductor with an impurity band (IB) that contains metallic ions  in two different valency states,
say  $Me^{2+}$ and $Me^{3+}$, energy of an impurity is affected by (re)distribution of $Me^{2+}$ and $Me^{3+}$ ions
in its vicinity (because of different interaction energy between the impurity and $Me^{2+}$/$Me^{3+}$ ion 
due to different charge, ionic radii, and magnetic moment) \cite{BaBa14}.  
Considering a hypothetical rearrangement process (RP) 
causing repetitious changes in distribution of $Me^{2+}$ and $Me^{3+}$ ions on metal ion positions
with a characteristic time constant, $t_{r}$, this RP unconditionally causes also repetitive changes of 
the energies of impurities with the same characteristic time constant \cite{BaBa14}. 
Therefore, the energy $E_i$ of the impurity $i$ in the impurity band 
is not constant in time, but varies  within the interval 		
					$E_{i, min} \leq E_i \leq E_{i, max}$ \cite{BaBa14}.
		\begin{figure}[!b]
			\center{
				\resizebox{1.00\columnwidth}{!}{%
  				\includegraphics{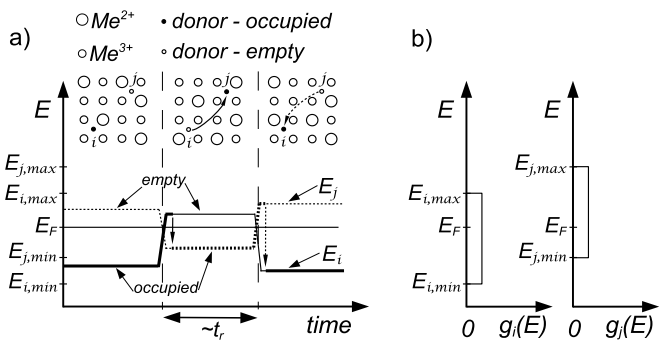}
        }
    	}
	 			\caption{Schematic depiction of  
time evolution of the energy of donor impurities ($i$, $j$)
with indicated electron hops
due to rearrangements of metallic ions (a) 
and time-averaged partial DOS 
corresponding to impurities $i$ and $j$ (b), as introduced formerly \cite{BaBa14}.
The inset in (a) shows examples of possible redistribution 
of $Me^{2+}$/$Me^{3+}$ ions and corresponding changes in occupation 
of donor sites.}
			\label{fig1}
		\end{figure}
%
%
%
Assuming that Fermi level,
$E_F$, lies in the IB formed by one type of impurities (e.g. donors)
and is characterized by a constant density of states (DOS),
	 $g(E)=g(E_F)$,
and supposing that dynamical changes of the impurity energies are adequately characterized by 
a typical width of the energy interval
$E_{i, max} - E_{i, min} = \Delta E_{i} \approx E_{0}$, 
then there must exist a subnetwork of impurities $i^*$ 
with energy $E_{i^{*}}$ satisfying the condition \cite{BaBa14}
\begin{equation}
				E_F - E_{0} < E_{i^{*}} < E_F + E_{0}.
				\label{Ef-band}
\end{equation}
The concentration of impurities in this subnetwork is  
$N^{*}= E_{0}g(E_F)$, 
thus a typical distance between two nearest impurities is 
$ R^{*}=[E_{0}g(E_F)]^{-1/3}$ \cite{BaBa14}.
The impurities defined by the inequality (\ref{Ef-band}) have a unique property: 
rearrangements of $Me^{2+}/Me^{3+}$ ions in the vicinity of these impurities cause
that some occupied impurity energy levels can shift  
from the region below $E_F$ to the region above $E_F$, and analogously,
some empty ones from the region above $E_F$ can shift under $E_F$ \cite{BaBa14}.
Such processes, driven by the RP create favorable conditions
for electron hops (tunneling) to empty sites of lower energy, 
i.e. hops with zero activation energy, as depicted in Fig. \ref{fig1}. 
It was shown  \cite{BaBa14} that probability of RP-induced hops,
$P^{*}_{rp}$, at sufficiently low temperatures can be expressed as  
%
$P^{*}_{rp} \propto  \nu^{*}_{rp} e^{-2\alpha R^{*}}$,
%
where $\alpha^{-1}$ is a localization length,  
and 
$\nu^{*}_{rp}$ is defined  as the time-averaged
probability per unit volume that for a {\em temporarily} occupied site belonging to the 
subnetwork defined by the inequality (\ref{Ef-band}) 
there will appear an empty state of lower energy in the subnetwork as a consequence of the RP. 
Accepting the fact that valence fluctuations (VFs) represent a special case of the RP,
probability of valence-fluctuation induced hop,
			 $P^{*}_{vf}$ can be expressed as 
\begin{equation}
P_{vf}^{*} \propto \nu_{vf}^{*} e^{-2\alpha R^{*}},
	\label{hopVF}
 \end{equation}		 
where $\nu_{vf}^{*}$ is defined analogously as $\nu_{rp}^{*}$ \cite{BaBa14}.
%
%
Direct consequence of Eq. (\ref{hopVF}) is that hopping probability 
exponentially increases with decrease of hop distance $R^{*}$, what implicates
an enhanced conductivity in the vicinity of the surface
because of higher concentration of lattice imperfections 
which can play role of hopping centers \cite{BaBa14}.

Now let us focus on another fundamental impact of the RP. 
Due to the RP energy levels of impurities defined by the inequality (\ref{Ef-band})
		repetitively pass through the Fermi level, resulting in possible 
		temperature non-activated electron hops, as depicted in Fig.~\ref{fig1}.
In the simplest case, 
when $|E_j-E_i|$ is of the order of the Debye energy or smaller,
and $kT$ is small compared to $|E_j-E_i|$,
the intrinsic transition rate $\gamma_{ij}$ for an electron hopping from a site 
		$i$ with energy $E_i$ to an empty site $j$ with energy $E_j$
		is well approximated by the ``quantum-limit'' hopping formula \cite{Ambegaokar71} 
\begin{eqnarray}
		\gamma_{ij} =	\gamma_{0}e^{-2\alpha R_{ij} - (E_j-E_i)/kT} \mbox{ for } E_j > E_i\\
		\gamma_{ij} = \gamma_{0}e^{-2\alpha R_{ij}} \mbox{ for } E_j < E_i ,
				\label{TNAH}
	\end{eqnarray} 
		where $k$ is Boltzmann constant, $R_{ij}$ is the distance between sites $i$ and $j$, 
		and $\gamma_{0}$ is some constant, which depends 
		on the electron-phonon coupling strength, the phonon density of states,
		and other properties of the material, but which depends only weakly
		on the energies $E_i$ and $E_j$ or on $R_{ij}$  \cite{Ambegaokar71}. 
According to Eq.~(4), the intrinsic transition rate of the electron hopping
to a site of less energy decreases exponentially with increasing $R_{ij}$.
Because $\gamma_{0}$ is finite,  $\gamma_{ij}$ must be also finite,
so there is always non-zero time interval until an electron can hop (tunnel)
to some empty site of less energy, while this time interval increases 
with increasing distance between sites.
Therefore, if the RP is a ground state  property of a material,  
        finite $\gamma_{ij}$ implicates a non-zero probability 
				of finding some occupied energy levels above $E_F$ 
				in the ground state.
Neglecting electron-electron interactions
except that not more than one electron can occupy a single site,
time averaged occupation number of site $i$ 
in thermal equilibrium
can be expressed in form $\left\langle n_i\right\rangle = 1/\left\{1 +\exp [(E_i - E_F)/kT]\right\} $ 
\cite{Ambegaokar71},  
what for $T = 0$~K unconditionally requires 
that all sites having energy below $E_F$ are occupied  
and all sites having energy above $E_F$ are empty. 
Therefore, if there exist some occupied sites with energy levels above $E_F$ 
in the ground state of a material with the RP, then the charge carrier subsystem 
in the ground state of this material can not be at absolute zero.

Because VFs can be regarded as a special case of the RP,
	analogous scenario as indicated above can be expected
in	valence fluctuating semiconductors, too. 
In principle, VFs are RPs with a short characteristic time constant $t_{r}$.
However, according to Heisenberg relation
 if $t_{r}$ is short enough, 
uncertainty of energy of the electron occupying site $i$ ($\Delta E_{i} \approx \hbar / t_{r}$)
may cause situation that
 energy of the electron occupying site $i$ 
becomes indistinguishable from the 
energy of empty site $j$, to which the electron "is going" to hop.
This limit case has no essential impact on the
valence-fluctuation induced hopping transport as proposed before \cite{BaBa14},
such as hops between sites $i$ and $j$ with indistinguishable energies 
can also be treated as hops/tunneling not requiring thermal activation, 
however, it becomes crucial in discussion on possible effect of VFs on the 
energy distribution function of localized charge carriers in the ground state 
of valence fluctuating materials, such as 
SmB$_6$. 
The charge fluctuation rate in SmB$_6$ estimated from
phonon spectroscopy studies  \cite{Zirngiebl86,Mock86}
is 200~cm$^{-1}$ $-$ 650~cm$^{-1}$, what 
corresponds to the characteristic time $\tau_{cf}$ (which defines time
changes of the impurity energy due to changes of 
the Coulomb interaction between the impurity 
and the surrounding Sm$^{2+}$/Sm$^{3+}$ ions) 
of $5.1 \times 10^{-14}$~s $-$ 1.7 $\times 10^{-13}$~s.
According to Heisenberg relation
	the uncertainty of the energy of electron occupying the impurity site can be estimated as 
	$\Delta E \approx \hbar / \tau_{cf}$, 
what  yields for  SmB$_6$ $\Delta E$ between 4~meV and 13~meV.
Because energies of the impurities  
fluctuating within the interval of 
$E_{i, max} - E_{i, min} \approx \Delta E_{0}$
lie in the impurity band of the width $W_{IB}$
that is located in the forbidden gap of the width $E_g$, 
 the inequality 
$\Delta E_{0} \leq W_{IB} < E_g$ is satisfied.
The value of $E_g$ 
detected by many experimental techniques is between 2 and 20~meV in SmB$_6$
\cite{Nickerson71,Allen79,Kasuya79,Frankowski82,Travaglini84,Ohta91,Namba93,Cooley95,Gorshunov99,Flachbart01,Gabani03,Batkova06,Batkova08},
		what can be an indication that the above estimated uncertainty 
		of the energy of electron occupying an impurity site 
		($\Delta E$)  is greater or comparable with
	the width of the impurity band ($W_{IB}$).
Based on this we suppose that the whole interval in which the energy of impurity fluctuates (because of fluctuations of local physical parameters which affect its energy) defines at the same time the energy uncertainty of electron occupying the impurity site (as a consequence of a characteristic time of fluctuations and uncertainty principle). 
Taking into account this supposition we present 
energy diagram model of impurity sites 
in SmB$_6$  as follows.  

\begin{figure}[!t ]
			\center{
				\resizebox{1.00\columnwidth}{!}{%
  				\includegraphics{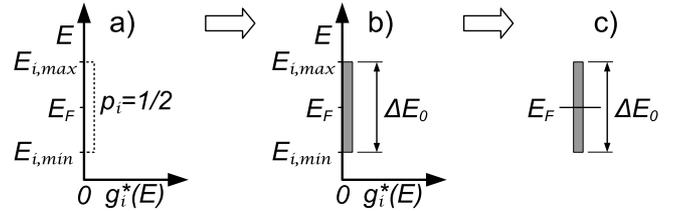}
        }
    	}
 			\caption{Schematic depiction of the UBIE which is in this example centered at the Fermi level.
The UBIE is characterized by the partial DOS, $g_{i}^{*}(E)$, 
and  by the time averaged probability $p_{i}$ of the occupation by an electron.
           The $p_{i}$ is indicated by number (a), or expressed using a color scale 
(black for $p_{i} = 1$, white for $p_{i} = 0$, 
		and corresponding percentage of gray for $0 < p_{i} < 1$) (b).
The same as in (a) and (b) can be expressed in the most simplified form shown in (c).}
		\label{fig2new}
		\end{figure}

Let us associate hopping sites in SmB$_6$ with donor-type impurity states
having fluctuating energy levels lying in the impurity band of SmB$_6$. 
Each hopping site is characterized by energy interval of the width $\Delta E_{0}$
defining energy broadening/uncertainty of this site.
Thus we treat the hopping center  
as a narrow band of the energy width $\Delta E_{0}$,
and we denote it as uncertainty band of the impurity energy (UBIE). 
The UBIE of a site $i$ is defined by the partial DOS, $g_{i}^{*}(E)$, 
			which is non-zero and constant within the 
				energy interval $\left\langle E_{i, min},E_{i, max} \right\rangle$
				and zero outside it, 
				and  by a time averaged probability of the occupation of site $i$ 
				by an electron, $ p_{i}\in \left\langle 0,1 \right\rangle $, 
				as depicted in Fig. \ref{fig2new}a. 
Expressing $p_{i}$ using a color scale
		the UBIE can be depicted as done in Fig.~\ref{fig2new}b, 
		or in the most simplified form in Fig.~\ref{fig2new}c.
%
%
%
%
%
%
	\begin{figure}[!b]
			\center{
				\resizebox{1.00\columnwidth}{!}{%
  				\includegraphics{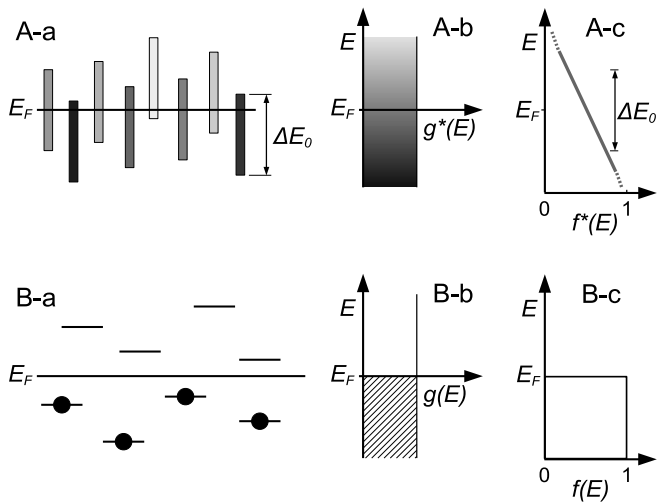}
        }
    	}
 			\caption{Schematic depiction of UBIEs located in the vicinity 
of Fermi level in the impurity band of SmB$_6$ (A-a) 
with the corresponding DOS diagram (A-b),  
and qualitative sketch of EDF of charge carriers (A-c)
when thermal reservoir is at $T=0$~K (see explanation in text).
For comparison, depiction of energies of donor levels 
in the IB of classical semiconductor 
with a corresponding DOS diagram (see e. g. \cite{Shklovskij84})
and EDF in the ground state is shown 
in (B-a), (B-b), and (B-c), respectively.}
		\label{fig2}
		\end{figure}
%
How the situation in the impurity band of  SmB$_6$ in very close vicinity of the Fermi level 
can look like show illustrations in  Fig.~\ref{fig2}A. 
Here the hopping sites are represented by UBIEs crossing the Fermi level ($0 < p_{i} < 1$),
and the partial DOS of the individual impurity, $g_{i}^{*}(E)$, is shown as a bar
		of corresponding dimensions and position in the energy scale (see Fig.~\ref{fig2}A-a).
We consider that all UBIEs have identical properties except their 
position in energy scale, and that their  distribution  in the 
impurity band of SmB$_6$ is uniform. 
Thus, the total DOS, $g^{*}(E)$, 
being a sum of all partial DOS of individual impurity states, $g_{i}^{*}(E)$,
is constant (see Fig.~\ref{fig2}A-b). 
The time averaged probability of the occupation of site $i$ 
				by an electron, $ p_{i} $, 	we define as the average value of the 
function $f_{0}(E_i,T) = 1/\left\{1 +\exp [(E_i - E_F)/kT]\right\}$ over the 
energy interval $\left\langle E_{i, min},E_{i, max} \right\rangle$.
At such conditions, 
$p_{i} = 1/2$ for the UBIE centered at the Fermi level 
and decreases/increases at increasing/decreasing $E_{i, min}$ (and $E_{i, max}$). 
The time averaged probability of occupation of the state 
at energy $E$ in the impurity band, 
 $f^{*}(E)$, can be determined  
    as average value of $p_{i}$ of all UBIEs with 
		 non-zero $g_{i}^{*}(E)$.
The behavior of	$f^{*}(E)$, qualitatively depicted in Fig.~\ref{fig2}A-c 
(and  also visualized by color representation in  Fig.~\ref{fig2}A-b)
 is a continuous function of energy in a close vicinity of the Fermi level,
	having a finite slope $\partial f^*(E)/\partial E$ at the Fermi level.
This "energy broadening" of $f^{*}(E)$ having its  origin in VFs cannot be less than broadening/uncertainty 
introduced by VFs (expressed by the width of UBIEs, $\Delta E_{0}$). 
An important point is that energy broadening induced by VFs 
will take place in SmB$_6$ also in the case if no thermal broadening can be expected
%
%
This is a fundamental difference in comparison to a 
classic semiconductor at $T=0$~K (see Fig.~\ref{fig2}B),
where all impurity energy levels 
below the Fermi level are occupied and all ones above the Fermi level 
are empty (as depicted in Fig.~\ref{fig2}B-a and Fig.~\ref{fig2}B-b),
so the energy distribution function, $f(E)$,  is a step function 
(as depicted in Fig.~\ref{fig2}B-c).

Now let us consider an imaginary experiment.
Let SmB$_6$ sample is in ideal thermal contact 
with a thermal reservoir at absolute zero, so the charge carrier subsystem of SmB$_6$ has to be in the ground state.
Supposing presence of VFs  in the ground state of SmB$_6$, 
the charge carrier subsystem in the ground state of SmB$_6$	
			will be characterized by 	monotonic and continuous function of energy	$f^{*}(E)$ in a close vicinity of the Fermi level,
		like shown in Fig.~\ref{fig2}A-c.  
If  energy distribution function of the charge carrier subsystem (being in thermal equilibrium, and in absence of electrical field)
can be approximated by Fermi-Dirac distribution function,  
 $f_0(E, T) = 1/\left\{1 +\exp [(E - E_F)/kT]\right\}$ at lowest temperatures,
then the energy distribution function of charge carriers in the ground state of SmB$_6$ 
can be approximated by $f^{*}(E) \approx 1/\left\{1 +\exp [(E - E_F)/kT_{CGS}]\right\}$,
where $T_{CGS}$ represents the temperature 
of the charge carrier subsystem in the ground state. 
Essential point is that $T_{CGS}$ has to be non-zero, 
such as zero value of $T_{CGS}$ would require that $f^{*}(E)$ is a step function,
what as follows from the discussion above cannot be the case in SmB$_6$.
It can be therefore concluded that temperature of  charge carrier subsystem 
in the ground state of SmB$_6$ cannot be zero, thus neither the electrical conductivity of SmB$_6$
in the ground state  can be zero,
such as zero electrical conductivity is the property
of a semiconductor with hopping conduction 
only at $T=0$~K. 
%
Consequently, although the SmB$_6$ 
sample is in ideal thermal contact 
with a thermal reservoir, which is at temperature $T << T_{CGS}$, 
 physical properties of SmB$_6$ that are governed by 
energy distribution function of charge carrier subsystem will
reveal only negligible dependence on the temperature of the thermal reservoir,
while a ''crossover'' to this regime will be observed at temperatures 
of thermal reservoir comparable with $T_{CGS}$.

With regard to the discussion in the paragraph above  it is worthy to mention 
that detailed resistivity studies of SmB$_6$
down to 15~mK reported by Kasuya and coworkers \cite{Kasuya79} 
have shown that the best fit of their experimental data down to 100~mK 
is represented by the formula
$\sigma(T)=7,45\times \exp{[-19,3/(T+4,87)]}$~[$\Omega^{-1}$cm$^{-1}$],
while the resistivity is nearly constant below 100~mK.
Surprisingly, such fit 
corresponds to Arrhenius-type activation process 
with ``shift" in temperature scale by 4,87~K.
If we associate this temperature shift with the non-zero temperature 
of charge carriers in the ground state of SmB$_6$, $T_{CGS}= 4,87$~K,
Kasuya's result can be interpreted by means that at cooling SmB$_6$ sample,
when temperature of the cooling system (thermal reservoir) 
approaches absolute zero the sample conductivity  approaches the value expected for Arrhenius-type activation process
at temperature of charge carriers of $4,87$~K.
In addition, such estimated $T_{CGS}$ yields surprisingly good correspondence 
with the  fact that plateau in the dc resistivity temperature dependence 
of SmB$_6$ samples of various qualities is observed below 5~K \cite{Allen79,Kasuya79,Cooley95,Gorshunov99,Batkova06,Wolgast13,Kim13,Phelan14,Chen15,Syers15,Nickerson71,Gabani03}. 

In association with impact of VFs on the ground state properties 
of charge carrier subsystem in SmB$_6$ it should be also emphasized 
that valence fluctuations of Sm ions are associated with fluctuations of 
their ionic radii,
consequently causing corresponding lattice vibrations. 
In a ''classic'' crystal lattice at absolute zero there are no phonons 
\cite{Srivastava90}, 
so here arises a question about energy distribution function of phonons, 
which assist to VFs in the ground state of SmB$_6$.
A non-zero thermal coupling between the lattice and charge carrier subsystem having 
a non-zero temperature in the ground state   
implicates a non-zero ground state temperature of the phonon subsystem, too. 
However, discussion of this issue is beyond the scope of this paper.

It can be concluded that discussed effect of valence fluctuations on the energy distribution function 
of localized charge carriers yields alternative explanation for the resistivity saturation in SmB$_6$ 
at lowest temperatures, while one does not exclude possibility of existence of metallic surface states 
in this material (e.g. topologically protected \cite{Dzero10, Dzero12, Lu13, Alexandrov13}
or polarity-driven "trivial" surface states \cite{Zhu13}).
Moreover, this effect allows us to explain also the ground state properties of nonstoichiometric SmB$_6$ samples. 
It would be worthy to emphasize here that it was recently suggested that the presence of even 1~\% of Sm vacancies
 leads to a smearing of the bulk hybridization gap in SmB$_6$ samples and would result in a breakdown of the KTI state \cite{Valentine16}. 
Thus, only the most stoichiometric SmB$_6$ samples possess a bulk gap necessary for the topological Kondo insulator state \cite{Valentine16}. 
However, recent electrical resistivity studies of vacant samples Sm$_{1-x}$B$_6$, for x ranging from $0.03$ to $0.2$, 
reveal resistivity saturation at lowest temperatures similar to the stoichiometric SmB$_6$ \cite{Pristas14,Gabani16}. 
Because concentration of Sm vacancies in these samples substantially exceeds level of 1~\%, 
topologically protected surface states should not be present there, what indicates a need for an alternative scenario to KTI 
allowing to  explain the resistivity saturation in such nonstoichiometric SmB$_6$ samples.

In fact, here discussed effect of valence fluctuations on the ground state properties of SmB$_6$ brings new light into the problem,
 and can help to resolve the puzzle, indicating following possibilities of realization of the ground state in SmB$_6$: 
(i) Considering "ideal quality" SmB$_6$ single crystals without an impurity band  
or highest quality single crystals with very small impurity band and Fermi level \emph{not} lying in the impurity band, 
the bulk should be insulating in the ground state (as there are no impurity states fluctuating around Fermi level, 
hich could be responsible for the phenomena discussed in this paper). 
However, because there are still present lattice imperfections in the near-surface region
(or at least unterminated Sm- and B- bonds on the surface) causing increase of density of states in the impurity band, 
Fermi level in the near-surface region is supposed to lie in the impurity band, 
thus the ground state energy distribution function of localized charge carriers in the near-surface region reveals energy 
"broadening", resembling effect of thermal broadening at non-zero temperature. 
This infers that near-surface region of highest-quality SmB$_6$ samples is electrically conductive in the ground state.
If, in addition, topologically protected surface states or/and polarity-driven "trivial" surface states are formed there,
 the (electrically conductive) near-surface region in the ground state of SmB$_6$ contains also a metallic region/surface. 
On the other hand, (ii) in "real world" high-quality SmB$_6$ single crystals with low concentration of lattice imperfections, 
the impurity band containing Fermi level is present in accordance with experimental observations. 
Thus, the energy distribution function of localized charge carriers in the near-surface region, as well as in the bulk,
reveals energy "broadening" in the ground state. 
This infers that the near-surface region, as well as bulk of the material, is electrically conductive in the ground state. 
If, moreover, topologically protected surface states or/and polarity-driven "trivial" surface states are formed there, 
the near-surface region of such "real world" high-quality SmB$_6$ crystals contain also a metallic (sub)region. 
Finally, (iii) in SmB$_6$ with reasonably high concentration of lattice imperfections, 
 the impurity band containing Fermi level with correspondingly higher DOS can be supposed to be present. 
Thus, the ground state energy distribution function of localized charge carriers in the near-surface region, 
as well as in the bulk, reveals energy "broadening", analogously as in the previous case. 
Therefore, the near-surface region, as well as the bulk of the material is expected to be electrically conductive 
in the ground state, while conductivity in the near-surface region is enhanced. 
 Deviation from the stoichiometry can cause that topologically protected surface states are not formed, 
what however still does not exclude possible presence of metallic surface states, e.g. polarity-driven "trivial" ones. 
If metallic surface states do not exist there, the near-surface region will just reveal enhanced conductivity, 
as it follows from the scenario of valence-fluctuation induced hopping transport \cite{BaBa14}.

Finally, we would like to 
emphasize that similar effects as described here for SmB$_6$ we expect also in other 
related materials with valence-fluctuating ground state, e.g. YbB$_{12}$.
According to our opinion, 
the absence of resistivity divergence at lowest temperatures 
is a fingerprint of this class of materials,
being at the same time a proof of valence fluctuating ground state. 
We believe that here proposed scenario  not only represents a base 
for  understanding the underlying physics  in valence fluctuating	 semiconducting  compounds at temperatures close to absolute zero, 
		but it also indicates a necessity to consider similar phenomena in many other materials 
		with ''dynamical ground state'', 
		especially those obeying physical properties which can not be adequately understood 
		presumably supposing the ground state being  at absolute zero.

This work was supported by 
the Slovak Scientific Agency VEGA (Grant No.~2/0015/17).


\begin{thebibliography}{10}
\expandafter\ifx\csname url\endcsname\relax
  \def\url#1{\texttt{#1}}\fi
\expandafter\ifx\csname urlprefix\endcsname\relax\def\urlprefix{URL }\fi

\bibitem[{Allen et~al.(1979)Allen, Batlogg, and Wachter}]{Allen79}
J.~W. Allen, B.~Batlogg, P.~Wachter, Large low-temperature {Hall} effect and
  resistivity in mixed-valent {SmB}$_{6}$, Phys. Rev. B 20 (1979) 4807--4813.

\bibitem[{Kasuya et~al.(1979)Kasuya, Takegahara, Fujita, Tanaka, and
  Bannai}]{Kasuya79}
T.~Kasuya, K.~Takegahara, T.~Fujita, T.~Tanaka, E.~Bannai, Valence fluctuating
  state in {SmB}$_{6}$, Le Journal de Physique Colloques 40~(C5) (1979)
  C5--308--C5--313.

\bibitem[{Cooley et~al.(1995)Cooley, Aronson, Fisk, and Canfield}]{Cooley95}
J.~C. Cooley, M.~C. Aronson, Z.~Fisk, P.~C. Canfield, {SmB}$_{6}$: Kondo
  insulator or exotic metal?, Phys. Rev. Lett. 74 (1995) 1629--1632.

\bibitem[{Batkova et~al.(2006)Batkova, Batko, Konovalova, Shitsevalova, and
  Paderno}]{Batkova06}
M.~Batkova, I.~Batko, E.~S. Konovalova, N.~Shitsevalova, Y.~Paderno, Gap
  properties of {SmB}$_{6}$ and {YbB}$_{12}$: Electrical resistivity and
  tunnelling spectroscopy studies, Physica B 378 - 380 (2006) 618.

\bibitem[{Ioffe and Regel(1960)}]{Ioffe60}
A.~Ioffe, A.~Regel, Non-crystalline, amorphous, and liquid electronic
  semiconductors, Progress in Semiconductors (1960) 237.

\bibitem[{Mott and Davis(1971)}]{Mott71}
N.~Mott, E.~Davis, Electronic Processes in Non-Crystalline Materials, Clarendon
  Press, 1971.

\bibitem[{Edwards et~al.(1998)Edwards, Johnston, Rao, Tunstall, and
  Hensel}]{Edwards98}
P.~P. Edwards, R.~L. Johnston, C.~N.~R. Rao, D.~P. Tunstall, F.~Hensel, The
  metal insulator transition: {A} perspective, Philosophical Transactions of
  the Royal Society of London A: Mathematical, Physical and Engineering
  Sciences 356~(1735) (1998) 5--22.

\bibitem[{Shklovskij and Efros(1984)}]{Shklovskij84}
B.~I. Shklovskij, A.~L. Efros, Electronic Properties of Doped Semiconductors,
  Springer Series in Solid State Sciences, 1984.

\bibitem[{Dzero et~al.(2010)Dzero, Sun, Galitski, and Coleman}]{Dzero10}
M.~Dzero, K.~Sun, V.~Galitski, P.~Coleman, Topological {Kondo} insulators,
  Phys. Rev. Lett. 104 (2010) 106408.

\bibitem[{Dzero et~al.(2012)Dzero, Sun, Coleman, and Galitski}]{Dzero12}
M.~Dzero, K.~Sun, P.~Coleman, V.~Galitski, Theory of topological {Kondo}
  insulators, Phys. Rev. B 85 (2012) 045130.

\bibitem[{Lu et~al.(2013)Lu, Zhao, Weng, Fang, and Dai}]{Lu13}
F.~Lu, J.~Zhao, H.~Weng, Z.~Fang, X.~Dai, Correlated topological insulators
  with mixed valence, Phys. Rev. Lett. 110 (2013) 096401.

\bibitem[{Alexandrov et~al.(2013)Alexandrov, Dzero, and Coleman}]{Alexandrov13}
V.~Alexandrov, M.~Dzero, P.~Coleman, Cubic topological {Kondo} insulators,
  Phys. Rev. Lett. 111 (2013) 226403.

\bibitem[{Zhu et~al.(2013)Zhu, Nicolaou, Levy, Butch, Syers, Wang, Paglione,
  Sawatzky, Elfimov, and Damascelli}]{Zhu13}
Z.-H. Zhu, A.~Nicolaou, G.~Levy, N.~P. Butch, P.~Syers, X.~F. Wang,
  J.~Paglione, G.~A. Sawatzky, I.~S. Elfimov, A.~Damascelli, Polarity-driven
  surface metallicity in {SmB}$_{6}$, Phys. Rev. Lett. 111 (2013) 216402.

\bibitem[{Batko and Batkova(2014)}]{BaBa14}
I.~Batko, M.~Batkova, {SmB}$_{6}$: Topological insulator or semiconductor with
  valence-fluctuation induced hopping transport?, Solid State Commun. 196
  (2014) 18.

\bibitem[{Wolgast et~al.(2013)Wolgast, Kurdak, Sun, Allen, Kim, and
  Fisk}]{Wolgast13}
S.~Wolgast, C.~Kurdak, K.~Sun, J.~W. Allen, D.-J. Kim, Z.~Fisk, Low-temperature
  surface conduction in the {Kondo} insulator {SmB}$_{6}$, Phys. Rev. B 88
  (2013) 180405.

\bibitem[{Kim et~al.(2013)Kim, Thomas, Grant, Botimer, Fisk, and Xia}]{Kim13}
D.~J. Kim, S.~Thomas, T.~Grant, J.~Botimer, Z.~Fisk, J.~Xia, Surface {Hall}
  effect and nonlocal transport in {SmB}$_{6}$: Evidence for surface
  conduction, Scientific Reports 3 (2013) 1.

\bibitem[{Syers et~al.(2015)Syers, Kim, Fuhrer, and Paglione}]{Syers15}
P.~Syers, D.~Kim, M.~S. Fuhrer, J.~Paglione, Tuning bulk and surface conduction
  in the proposed topological {Kondo} insulator {SmB}$_{6}$, Phys. Rev. Lett.
  114 (2015) 096601.

\bibitem[{Jiang et~al.(2013)Jiang, Zhang, Sun, Chen, Yev, Xu, Ge, Tan, Niu,
  Xia, Xie, Li, Chen, Wen, and Feng}]{Jiang13}
S.~L. Jiang, T.~Zhang, Z.~Sun, F.~Chen, Z.~Yev, M.~Xu, Q.~Ge, S.~Tan, X.~Niu,
  M.~Xia, B.~Xie, Y.~Li, X.~Chen, H.~Wen, D.~Feng, Observation of possible
  topological in-gap surface states in the {Kondo} insulator {SmB}$_{6}$ by
  photoemission, Nature Communications 4 (2013) 3010.

\bibitem[{Neupane et~al.(2013)Neupane, Alidoust, Xu, Kondo, Ishida, Kim, Liu,
  Belopolski, Jo, Chang, Jeng, Durakiewicz, Balicas, Lin, Bansil, Shin, Fisk,
  and Hasan}]{Neupane13}
M.~Neupane, N.~Alidoust, S.-Y. Xu, T.~Kondo, Y.~Ishida, D.~J. Kim, C.~Liu,
  I.~Belopolski, Y.~J. Jo, T.-R. Chang, H.-T. Jeng, T.~Durakiewicz, L.~Balicas,
  H.~Lin, A.~Bansil, S.~Shin, Z.~Fisk, M.~Z. Hasan, Surface electronic
  structure of the topological {Kondo}-insulator candidate correlated electron
  system {SmB}$_{6}$, Nature Communications 4 (2013) 2991.

\bibitem[{Kim et~al.(2014)Kim, Xia, and Fisk}]{Kim14}
D.~J. Kim, J.~Xia, Z.~Fisk, Topological surface state in the {Kondo} insulator
  samarium hexaboride, Nature Materials 13 (2014) 466�470.

\bibitem[{Li et~al.(2014)Li, Xiang, Yu, Asaba, Lawson, Cai, Tinsman, Berkley,
  Wolgast, Eo, Kim, Kurdak, Allen, Sun, Chen, Wang, Fisk, and Li}]{Li14}
G.~Li, Z.~Xiang, F.~Yu, T.~Asaba, B.~Lawson, P.~Cai, C.~Tinsman, A.~Berkley,
  S.~Wolgast, Y.~S. Eo, D.-J. Kim, C.~Kurdak, J.~W. Allen, K.~Sun, X.~H. Chen,
  Y.~Y. Wang, Z.~Fisk, L.~Li, Two-dimensional {Fermi} surfaces in {Kondo}
  insulator {SmB}$_{6}$, Science 346~(6214) (2014) 1208--1212.

\bibitem[{Ruan et~al.(2014)Ruan, Ye, Guo, Chen, Chen, Zhang, and Wang}]{Ruan14}
W.~Ruan, C.~Ye, M.~Guo, F.~Chen, X.~Chen, G.-M. Zhang, Y.~Wang, Emergence of a
  coherent in-gap state in the {SmB}$_{6}$ {Kondo} insulator revealed by
  scanning tunneling spectroscopy, Phys. Rev. Lett. 112 (2014) 136401.

\bibitem[{Phelan et~al.(2014)Phelan, Koohpayeh, Cottingham, Freeland, Leiner,
  Broholm, and McQueen}]{Phelan14}
W.~A. Phelan, S.~M. Koohpayeh, P.~Cottingham, J.~W. Freeland, J.~C. Leiner,
  C.~L. Broholm, T.~M. McQueen, Correlation between bulk thermodynamic
  measurements and the low-temperature-resistance plateau in {SmB}$_{6}$, Phys.
  Rev. X 4 (2014) 031012.

\bibitem[{Nakajima et~al.(2016)Nakajima, Syers, Wang, Wang, and
  Paglione}]{Nakajima15}
Y.~Nakajima, P.~Syers, X.~Wang, R.~Wang, J.~Paglione, One-dimensional edge
  state transport in a topological {Kondo} insulator, Nature Physics 12 (2016)
  213�217.

\bibitem[{Chen et~al.(2015)Chen, Shang, Jin, Zhao, Wu, Xiang, Xia, Wang, Luo,
  Wu, and Chen}]{Chen15}
F.~Chen, C.~Shang, Z.~Jin, D.~Zhao, Y.~P. Wu, Z.~J. Xiang, Z.~C. Xia, A.~F.
  Wang, X.~G. Luo, T.~Wu, X.~H. Chen, Magnetoresistance evidence of a surface
  state and a field-dependent insulating state in the {Kondo} insulator
  {SmB}$_{6}$, Phys. Rev. B 91 (2015) 205133.

\bibitem[{Hlawenka et~al.(2015)Hlawenka, Siemensmeyer, Weschke, Varykhalov,
  Sanchez-Barriga, Shitsevalova, Dukhnenko, Filipov, Gab\'ani, Flachbart,
  Rader, and Rienks}]{Hlawenka15}
P.~Hlawenka, K.~Siemensmeyer, E.~Weschke, A.~Varykhalov, J.~Sanchez-Barriga,
  N.~Y. Shitsevalova, A.~V. Dukhnenko, V.~B. Filipov, S.~Gab\'ani,
  K.~Flachbart, O.~Rader, E.~D.~L. Rienks, Samarium hexaboride: A trivial
  surface conductor, arXiv:1502.01542.

\bibitem[{Valentine et~al.(2016)Valentine, Koohpayeh, Phelan, McQueen, Rosa,
  Fisk, and Drichko}]{Valentine16}
M.~E. Valentine, S.~Koohpayeh, W.~A. Phelan, T.~M. McQueen, P.~F.~S. Rosa,
  Z.~Fisk, N.~Drichko, Breakdown of the {Kondo} insulating state in {SmB}$_{6}$
  by introducing {Sm} vacancies, Phys. Rev. B 94 (2016) 075102.

\bibitem[{Ambegaokar et~al.(1971)Ambegaokar, Halperin, and
  Langer}]{Ambegaokar71}
V.~Ambegaokar, B.~I. Halperin, J.~S. Langer, Hopping conductivity in disordered
  systems, Phys. Rev. B 4 (1971) 2612--2620.

\bibitem[{Zirngiebl et~al.(1986)Zirngiebl, Blumenroder, Mock, and
  Guntherodt}]{Zirngiebl86}
E.~Zirngiebl, S.~Blumenroder, R.~Mock, G.~Guntherodt, Relation of phonon
  anomalies to charge fluctuation rates in intermediate valence compounds,
  Journal of Magnetism and Magnetic Materials 54 (1986) 359 -- 360.

\bibitem[{Mock et~al.(1986)Mock, Zirngiebl, Hillebrands, G\"untherodt, and
  Holtzberg}]{Mock86}
R.~Mock, E.~Zirngiebl, B.~Hillebrands, G.~G\"untherodt, F.~Holtzberg,
  Experimental identification of charge relaxation rates in
  intermediate-valence compounds by phonon spectroscopy, Phys. Rev. Lett. 57
  (1986) 1040--1043.

\bibitem[{Nickerson et~al.(1971)Nickerson, White, Lee, Bachmann, Geballe, and
  Hull}]{Nickerson71}
J.~C. Nickerson, R.~M. White, K.~N. Lee, R.~Bachmann, T.~H. Geballe, G.~W.
  Hull, Physical properties of {SmB}$_{6}$, Phys. Rev. B 3 (1971) 2030 -- 2042.

\bibitem[{Frankowski and Wachter(1982)}]{Frankowski82}
L.~Frankowski, P.~Wachter, Point-contact spectroscopy on {SmB}$_{6}$, {TmSe},
  {LaB}$_{6}$ and {LaSe}, Solid State Commun. 41 (1982) 577.

\bibitem[{Travaglini and Wachter(1984)}]{Travaglini84}
G.~Travaglini, P.~Wachter, Intermediate-valent {SmB}$_{6}$ and the
  hybridization model: {An} optical study, Phys. Rev. B 29 (1984) 893--898.

\bibitem[{Ohta et~al.(1991)Ohta, Tanaka, Motokawa, Kunii, and Kasuya}]{Ohta91}
H.~Ohta, R.~Tanaka, M.~Motokawa, S.~Kunii, T.~Kasuya, Far-infrared transmission
  spectra of {SmB}$_{6}$, Journal of the Physical Society of Japan 60~(4)
  (1991) 1361--1364.

\bibitem[{Namba et~al.(1993)Namba, Ohta, Motokawa, Kimura, Kunii, and
  Kasuya}]{Namba93}
T.~Namba, H.~Ohta, M.~Motokawa, S.~Kimura, S.~Kunii, T.~Kasuya, Physica B
  186-188 (1993) 440.

\bibitem[{Gorshunov et~al.(1999)Gorshunov, Sluchanko, Volkov, Dressel, Knebel,
  Loidl, and Kunii}]{Gorshunov99}
B.~Gorshunov, N.~Sluchanko, A.~Volkov, M.~Dressel, G.~Knebel, A.~Loidl,
  S.~Kunii, Low-energy electrodynamics of {SmB}$_{6}$, Phys. Rev. B 59 (1999)
  1808--1814.

\bibitem[{Flachbart et~al.(2001)Flachbart, Gloos, Konovalova, Paderno,
  Reiffers, Samuely, and \v{S}vec}]{Flachbart01}
K.~Flachbart, K.~Gloos, E.~Konovalova, Y.~Paderno, M.~Reiffers, P.~Samuely,
  P.~\v{S}vec, Energy gap of intermediate-valent {SmB}$_{6}$ studied by
  point-contact spectroscopy, Phys. Rev. B 64 (2001) 085104.

\bibitem[{Gab\'ani et~al.(2003)Gab\'ani, Bauer, Berger, Flachbart, Paderno,
  Paul, Pavl\'{\i}k, and Shitsevalova}]{Gabani03}
S.~Gab\'ani, E.~Bauer, S.~Berger, K.~Flachbart, Y.~Paderno, C.~Paul,
  V.~Pavl\'{\i}k, N.~Shitsevalova, Pressure-induced {Fermi-liquid} behavior in
  the {Kondo} insulator {SmB}$_{6}$: Possible transition through a quantum
  critical point, Phys. Rev. B 67 (2003) 172406.

\bibitem[{Batkova et~al.(2008)Batkova, Batko, Konovalova, Shitsevalova, and
  Paderno}]{Batkova08}
M.~Batkova, I.~Batko, E.~S. Konovalova, N.~Shitsevalova, Y.~Paderno, Tunneling
  spectroscopy studies of {SmB}$_{6}$ and {YbB}$_{12}$, Acta Physica Polonica A
  113 (2008) 255.

\bibitem[{Srivastava(1990)}]{Srivastava90}
G.~Srivastava, The Physics of Phonons, CRC Press, 1990.

\bibitem[{Prist\'a\v{s} et~al.(2014)Prist\'a\v{s}, Gab\'ani, Flachbart,
  Filipov, and Shitsevalova}]{Pristas14}
G.~Prist\'a\v{s}, S.~Gab\'ani, K.~Flachbart, V.~Filipov, N.~Shitsevalova,
  Investigation of the energy gap in {Sm}$_{1-x}${B}$_{6}$ and
  {Sm}$_{1-x}${La}$_{x}${B}$_{6}$ {Kondo} insulators, Proceedings of the
  International Conference on Strongly Correlated Electron Systems (SCES2013) 3
  (2014) 012021.

\bibitem[{Gab\'ani et~al.(2016)Gab\'ani, Orend\'a\v{c}, Prist\'a\v{s},
  Ga\v{z}o, Diko, Piovar\v{c}i, Glushkov, Sluchanko, Levchenko, Shitsevalova,
  and Flachbart}]{Gabani16}
S.~Gab\'ani, M.~Orend\'a\v{c}, G.~Prist\'a\v{s}, E.~Ga\v{z}o, P.~Diko,
  S.~Piovar\v{c}i, V.~Glushkov, N.~Sluchanko, A.~Levchenko, N.~Shitsevalova,
  K.~Flachbart, Transport properties of variously doped {SmB}$_{6}$,
  Philosophical Magazine 96~(31) (2016) 3274--3283.

\end{thebibliography}

\end{document}